\documentclass[11pt]{article}
\usepackage[margin=1in]{geometry}
\usepackage{graphicx}
\usepackage{subcaption}
\usepackage{xcolor}

\def\beq{\begin{equation}}
\def\eeq#1{\label{#1}\end{equation}}
\def\eeqn{\end{equation}}

\def\beqa{\begin{eqnarray}}
\def\eeqa#1{\label{#1}\end{eqnarray}}
\def\eeqan{\end{eqnarray}}

\let\bar=\overbar

\def\Dslash{\not{\hbox{\kern-4pt $D$}}}
\def\dslash{\not{\hbox{\kern-2pt $\del$}}}

\def\msb{{\bar{\ssstyle M \kern -1pt S}}}

\def\Title#1{\begin{center} {\Large {\bf #1} } \end{center}}
\def\Author#1{\begin{center} {\normalsize {\sc #1} } \end{center}}
\def\Institution#1{\begin{center} {\normalsize {\it #1} } \end{center}}
\def\Abstract#1{\noindent {\normalsize {\bf Abstract:} {\normalfont #1}}}
\def\Conference{\vspace{4mm}\begin{raggedright} {\normalsize {\it Talk presented at the 2019 Meeting of the Division of Particles and Fields of the American Physical Society (DPF2019), July 29--August 2, 2019, Northeastern University, Boston, C1907293.} } \end{raggedright}\vspace{4mm}}

\begin{document}

%
%

\Title{Exotic Charmonium at BESIII}

\Author{Ilaria Balossino \\ on behalf of the BESIII Collaboration}

\Institution{Istituto Nazionale di Fisica Nucleare - Sezione di Ferrara, Ferrara, ITALY\\
			Institute of High Energy Physics, Chinese Academy of Science, Beijing, PRC	}

\Abstract{The BESIII Experiment at the Beijing Electron Positron Collider (BEPCII) collected large data samples for electron-positron collisions with center-of-mass above 4 GeV. The analysis of these samples has resulted in a number of surprising discoveries, such as the discoveries of the electrically charged $\mathrm{Z_c}$ structures, which, if resonant, cannot be accommodated in the traditional charm quark and anti-charm quark picture of charmonium. In this talk, we will review the current status of the analyses of the exotic states, as well as a number of other interesting features in the new BESIII data samples.}

\Conference

%
%
{\parindent0pt 

\section{Introduction}

The BESIII (BEijing Spectrometer III) Experiment is hosted at the Institute of High Energy Physics of Beijing, PRC. Since 2008, it works with a central detector installed at the Beijing Electron Position Collider II (BEPCII) optimized for flavour physics \cite{Ablikim:2009aa}. The whole environment is optimized for the so-called $\tau - charm$ energy regime between $2$ and $4.6~\mathrm{GeV}$ and allows to tune the center of mass energy on the point of interest. The design luminosity is $\mathit{L} = 10^{33}~\mathrm{cm^{-2}s^{-1}}$ at $3.77~\mathrm{GeV}$ and it has been achieved in 2016. \\
The detector is divided in a barrel and two end caps regions to cover the $93\%$ of $4 \mathrm{\pi}$ acceptance. The sub-detectors system is designed with a multi-layered drift chamber for momentum measurements of charge particles, a time-of-flight system for the identification of charged particles, and electromagnetic calorimeter to provide the trigger and measure the photons' energies and a muon system made of resistive plate chambers.\\
During the past years, BESIII has collected over $12~\mathrm{fb^{-1}}$ of data dedicated to charmonium-like states between  $ 4.18$ and $4.6~\mathrm{GeV}$. One of the goals of this experiment is to contribute to deepen the knowledge of the so-called XYZ states, exotic states that do not fit the standard quark model template and have unclear nature. The unique features of BESIII and BEPCII allows to measure with unprecedented precision the cross sections of very different final states and extract the different lineshape. In particular the possibility to collect large samples of data at the production cross sections helps to study the possible realation between those different states.
\\
Since the first discovery of a charmonium-like peak X(3872) made by BELLE in 2003 \cite{Choi:2003ue}, the hadronic states family called XYZ states placed in the energy region above the $DD^*$ threshold is expanding and challenging the theory for explaining their properties and behaviour. They do not behave like conventional states because they are either in over-abundance with respect to the predicted one (Y states) or they have non-zero electrical charge (Z states). X states, instead, are all the other unpredicted ones. 
The theory have different hypotesis that includes multi-quark states, meson molecules, hybrids, and hadron-quarkonium states \cite{Voloshin:2019rfo}.\\
As already pointed out, BESIII has the chance to tune the center of mass energy around the peak of the charmonium-like state and study the exotics states with more statistic and more precision.

\section{XYZ States at BESIII}

Here the most relevant results are presented among the different works that the collaboration is performing in this subject. First, an overview on the $X(3872)$ will be presented, followed by the results on the Z states, and then the proceeding will close with the Y states.

\subsection*{X(3872)}
\subsubsection*{$X(3872) \rightarrow \omega J/\psi$}
The resonant $X(3872)$ state, observed in 2003 by BELLE, and confirmed later by different experiments, still does not have a clear explanation. Experimentally is defined as an unconventional meson candidate. Since its proximity to the $\bar{D}^0D^{*0}$, theoretically could be interpreted as an hadronic molecule. This latter model predicts the decay $X(3872) \rightarrow \omega J/\psi$ could give the chance to look at its internal structure with a precise measurement of the decay rate that could allow to determine the ratio of the components that contribute to the wave function. Moreover, this process conserve isospin symmetry and so it could be the perfect probe. \\
With BESIII data samples is possible to investigate the process $e^+ e^- \rightarrow \gamma \omega J/\psi$ and therefore to search, in the $\omega J/\psi$ system, for the $X(3872)$ and study the $\sqrt{s}-$dependent production cross section, $\sigma[e^+ e^- \rightarrow \gamma X(3872)]$ \cite{Ablikim:2019zio}.
Such observations could help to undestand its nature and its connection with $Y(4260)$ and $Z_c(3900)$ states and resonances, and hit towars common underlying nature for them.\\
The $11~\mathrm{fb^{-1}}$ data allowed, for the first time, to observe with more than $5\sigma$ significance the $X(3872) \rightarrow \omega J/\psi$ decay. The $X(3872)$ mass extracted from the fit with a fixed width is $ m_{(X3872)} = 3873.3 \pm 1.1 \pm 1.0~\mathrm{MeV/c^2}$. However, to fit the full spectrum shown in Fig~\ref{fig:fig1}, it was needed and additional structure above $3.9~\mathrm{GeV}$. Two solutions are found for the structure: both statistically equivalent, they are compatible with a $X(3915)$ state, observed earlier by \cite{Abe:2004zs,Uehara:2009tx}.\\

\begin{figure}[htb]
\centering
\includegraphics[height=1.22in]{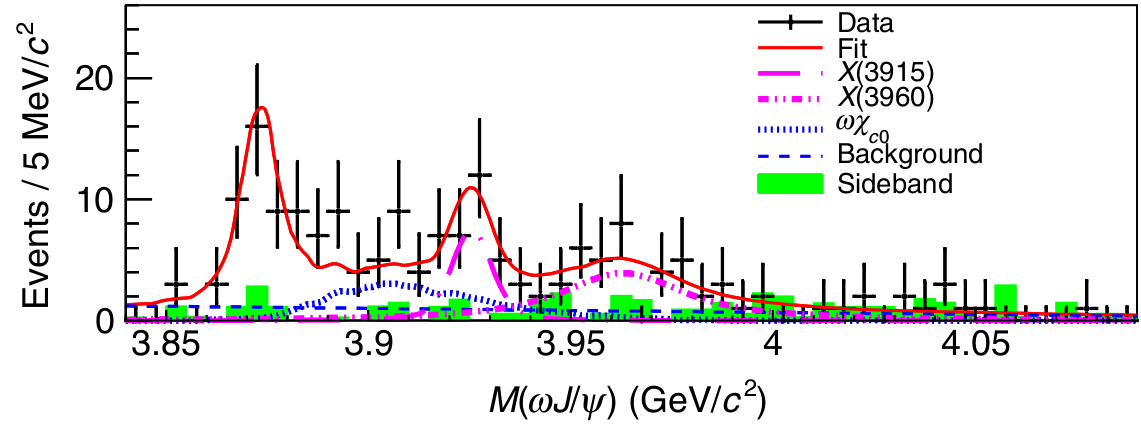}
\includegraphics[height=1.22in]{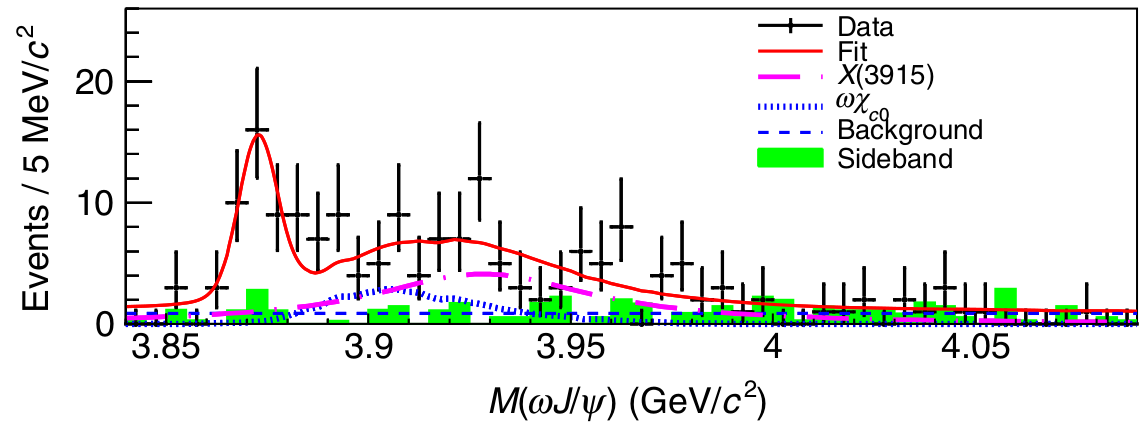}
\caption{$ \omega J/\psi$ invariant mass. Fit performed with $X(3872)$ plus an additional state: left - narrow $X(3915)$; right - broad $X(3915)$}
\label{fig:fig1}
\end{figure}

The studies on the production cross section of the process $e^+ e^- \rightarrow \gamma X(3872)$, performed in a center of mass energy range between $4$ and $4.6~\mathrm{GeV}$, are compared with the results of the process $e^+ e^- \rightarrow \gamma \pi\pi J/\psi$ to look for a transition like $Y(4220) \rightarrow \gamma X(3872)$ and study the isospin breaking contribute. A simultaneous Breit-Wigner fit is performed on the data shown in Fig~\ref{fig:fig2}. The extracted structure have a mass $M = (4200.6^{+7.9}_{-13.3} \pm 3.0)~\mathrm{MeV/c^2}$ and a width $\Gamma = (115^{+38}_{-26} \pm 12)~\mathrm{MeV}$. This result is compatible both with the recent parametrizations of $Y(4220)$ or $\psi(4150)$.\\
Moreover, the ratio extracted from the fits is  $R \left[ \frac{B(X(3872) \rightarrow \omega J/ \psi)}{B(X(3872) \rightarrow \pi^{+} \pi^{-} J/ \psi)} \right] = 1.6^{+0.4}_{-0.3} \pm 0.2$, which could provide important input for the hadronic molecule interpretation for the $X(3872)$ resonance.

\begin{figure}[htb]
\centering
\includegraphics[height=2in]{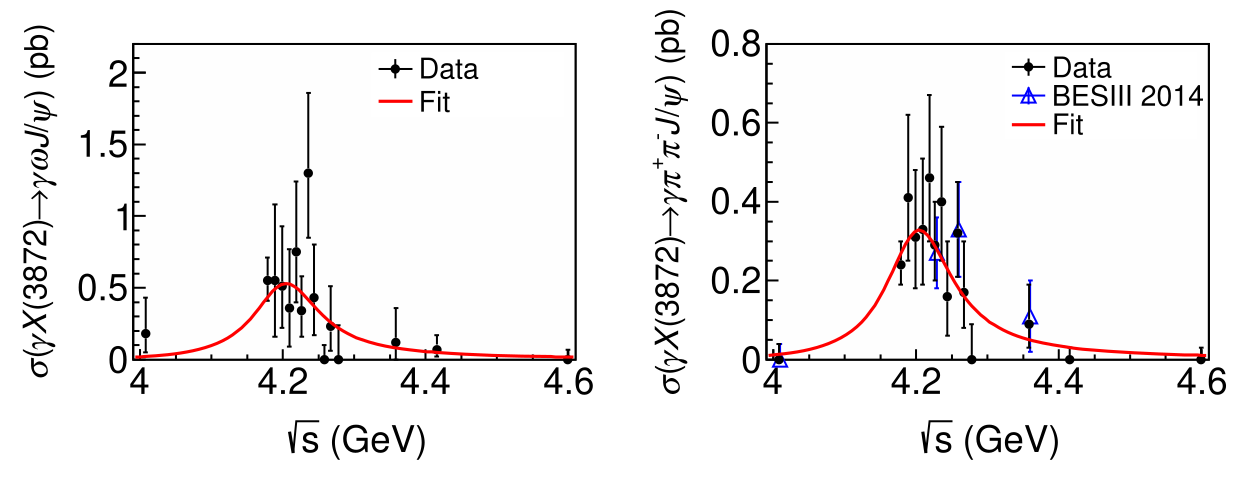}
\caption{Left - $e^+ e^- \rightarrow \gamma \omega J/\psi$ cross section lineshape. Right - $e^+ e^- \rightarrow \gamma \pi\pi J/\psi$ cross section lineshape. }
\label{fig:fig2}
\end{figure}

\subsubsection*{$X(3872) \rightarrow \pi^0 \chi_{cJ}(1P)$}
The study of the pionic transitions of the $X(3872)$ to the $\chi_{cJ}$ could help to clarify the nature of this states and point out to a specific model: a small transition would be a sign of a conventional $c\bar{c}$ state while a big one would support the tetraquark or molecular state hypotesis \cite{Dubynskiy:2007tj}.
This investigation involves the process $e^+ e^- \rightarrow \gamma X(3872)$ with $X(3872) \rightarrow \pi^0 \chi_{cJ} (J=0,1,2)$ analysing $9~\mathrm{fb^{-1}}$ of data with $4.15 < \sqrt{s} < 4.30 ~\mathrm{GeV}$ \cite{Ablikim:2019soz}. \\
Fig~\ref{fig:fig3} shows the first observation of the decay $X(3872) \rightarrow \pi^0 \chi_{c1}$ with a significance of $5.2\sigma$. Since no signal is present in the others energy regions, this result is compatible with a possible transition between this states and $Y(4220)$. The observation of a transition with such large branching ratio disfavour the interpretation of $X(3872)$ as a conventional $\chi_{c1}(2P)$ charmonium.

\begin{figure}[htb]
\centering
\includegraphics[height=2in]{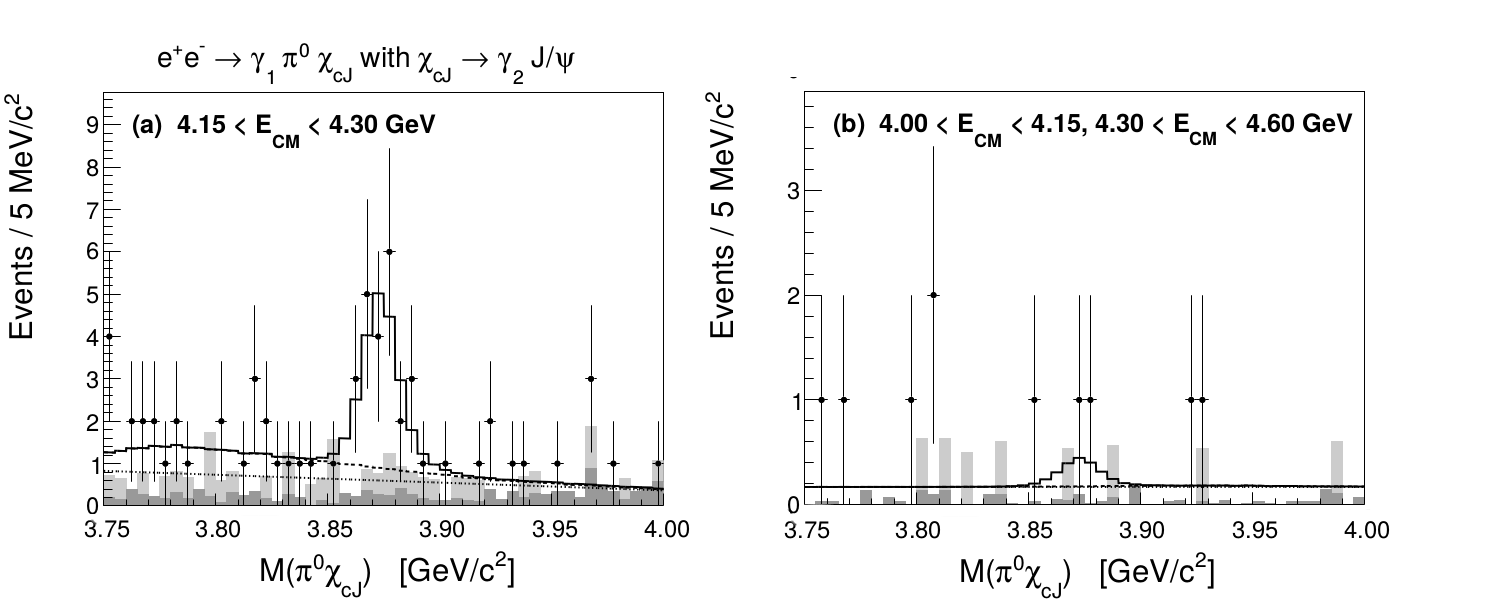}
\caption{$\pi^0 \chi_{cJ}$ invariant mass.}
\label{fig:fig3}
\end{figure}

\subsection*{Z}
\subsubsection*{$Z_{c}^{(')} \rightarrow \rho^{\pm} \eta_{c}$}

As for the other XYZ states, the nature of the $Z_c(3900)$ and $Z^{'}_c(4020)$ is not clear. The most frequent interpretations are either the compact tetraquark or the molecule ones. 
The discrimination between these two hypoteses could be given by the comparing the branching ratios of the decays into $\rho^{\pm}\eta_c$, $\pi^{\pm} J/\psi$ and $\pi^{\pm} h_c$.\\
The BESIII world largest data samples at $4.23,~4.26$ and $4.36~\mathrm{GeV}$ allowed to extract the preliminary results shown in Fig~\ref{fig:fig4} \cite{Ablikim:2019ipd}. Here an evidence of $e^+ e^- \rightarrow \pi Z_c, ~Z_c \rightarrow \rho \eta_c$ at $4.23 ~\mathrm{GeV}$ is present, but there is no sing of the process $e^+ e^- \rightarrow \pi Z^{'}_c, ~Z^{'}_c \rightarrow \rho \eta_c$ in any of the datasets.\\
The results found until now at $4.23 ~\mathrm{GeV}$ of the ratios are $R_Z \left[ \frac{\textit{B}(Z_c(3900)^{\pm} \rightarrow \rho^{\pm}\eta_c)}{\textit{B}(Z_c(3900)^{\pm} \rightarrow \pi^{\pm} J/\psi)} \right] = 2.2 \pm 0.9$ and $R_{Z'} \left[ \frac{\textit{B}(Z_c(4020)^{\pm} \rightarrow \rho^{\pm}\eta_c)}{\textit{B}(Z_c(4020)^{\pm} \rightarrow \pi^{\pm} h_c)} \right] < 1.6 $. They do not help the discrimination between the molecular and the tetraquark assignment \cite{Esposito:2014hsa}.\\

\begin{figure}[htb] 
\centering
\includegraphics[height=1.7in]{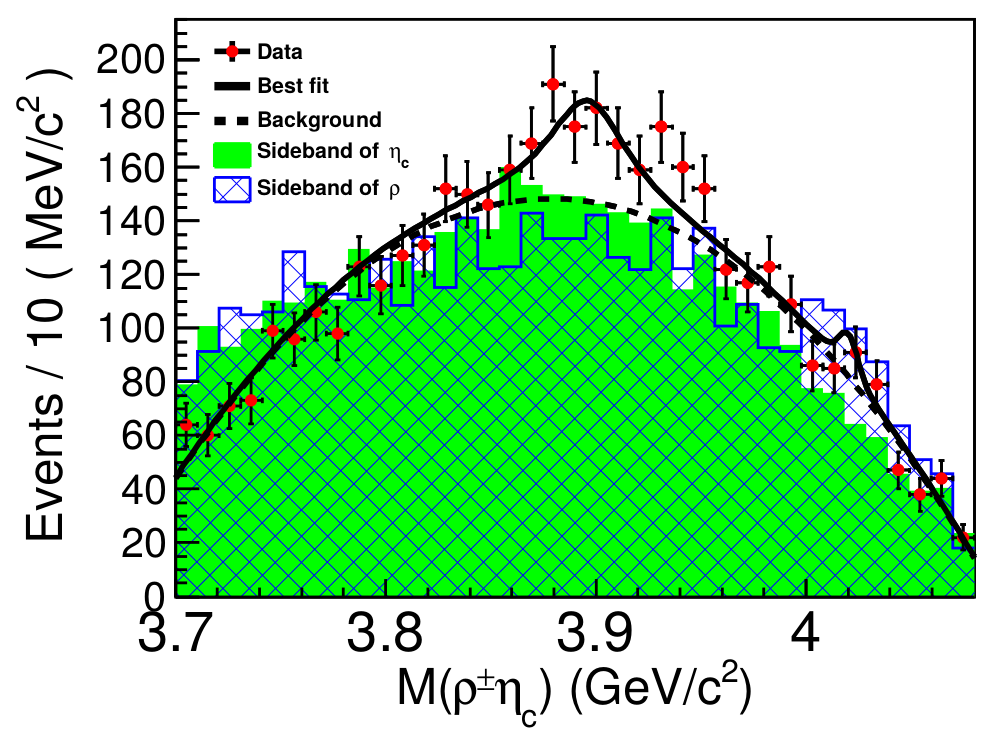}
\caption{Recoil mass against $\pi^{\pm}$ in $e^+ e^- \rightarrow \pi \rho \eta_c$.}
\label{fig:fig4}
\end{figure}

\subsection*{Y}
\subsubsection*{$Y(4220) \rightarrow \omega \chi_{c0}$}
Several Y states have been found in different experiment at different center-of-mass energy regions. Those states, as exotics states, do not have clear nature and could be hybrid, tetraquark or molecules states. BESIII contributed with their investigation thanks to the possibility to tune the beam energy at the point of interest.  The recent data collected at BESIII at nine energy points in the range $4.178 \leq \sqrt{s} \leq 4.278 ~\mathrm{GeV}$ allowed to study the process $e^+ e^- \rightarrow \omega \chi_{c0}$ ($\omega \rightarrow \pi^+ \pi^- \pi^0$; $\chi_{c0} \rightarrow \pi^+ \pi^- / K^+ K^-$) \cite{Ablikim:2019apl}.\\
This helped to confirm earlier observation as it is possible to see in Fig~\ref{fig:fig5}. Here a single resonance fit determine the mass to be $M = (4218.5 \pm 1.6\mathrm{(stat.)} \pm 4.0\mathrm{(sys.)})~\mathrm{MeV/c^2}$ and $\Gamma = (28.2 \pm 3.9\mathrm{(stat.)} \pm 1.6\mathrm{(stat.)})~\mathrm{MeV}$ compatible with the $Y(4220)$. \\
This study also included the angolar distribution of this process that underlined the contribution of a combination of a S and D-wave in the decay.\\
With respect to the present overview of this state, this results is consistent with the others for the mass, but not for the width (Fig~\ref{fig:fig6}). It is still not possible to conclude which is the nature of this state: higher statistic is needed for a more reliable conclusion.

\begin{figure}[htb] 
\begin{subfigure}[b]{0.5\linewidth}
\includegraphics[height=1.7in]{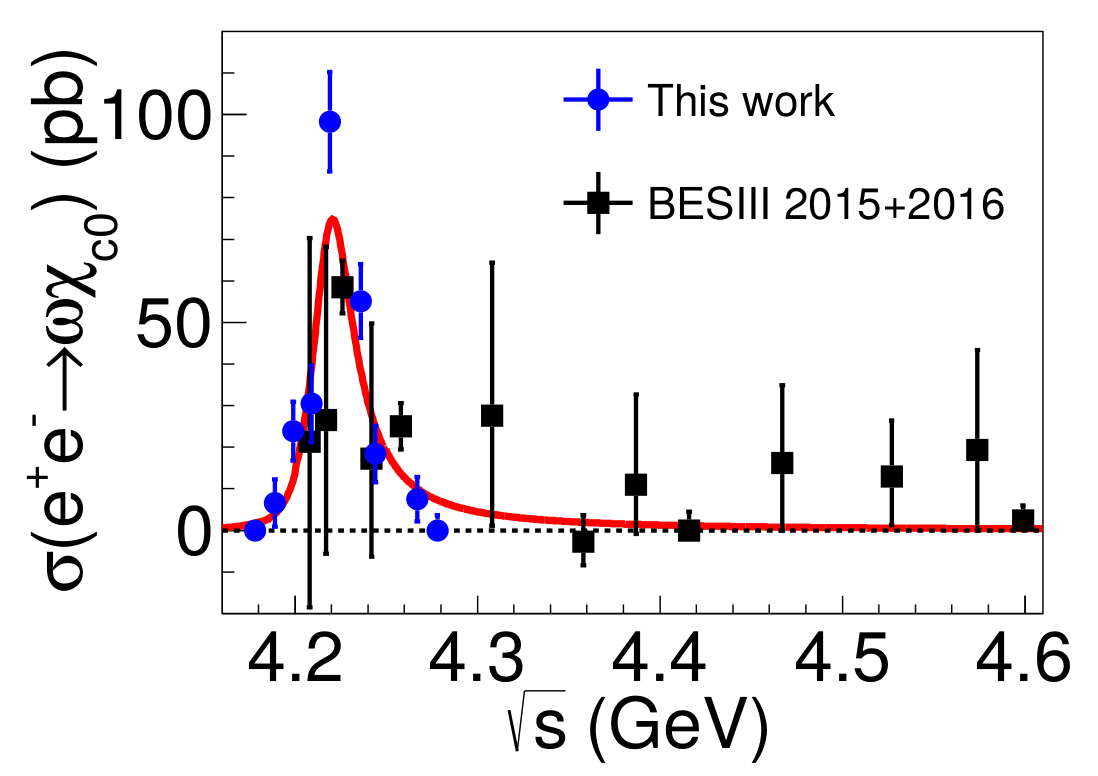}
\caption{$e^+ e^- \rightarrow \omega \chi_{c0}$.}
\label{fig:fig5}
\end{subfigure}
\begin{subfigure}[b]{0.5\linewidth} 
\includegraphics[height=1.7in]{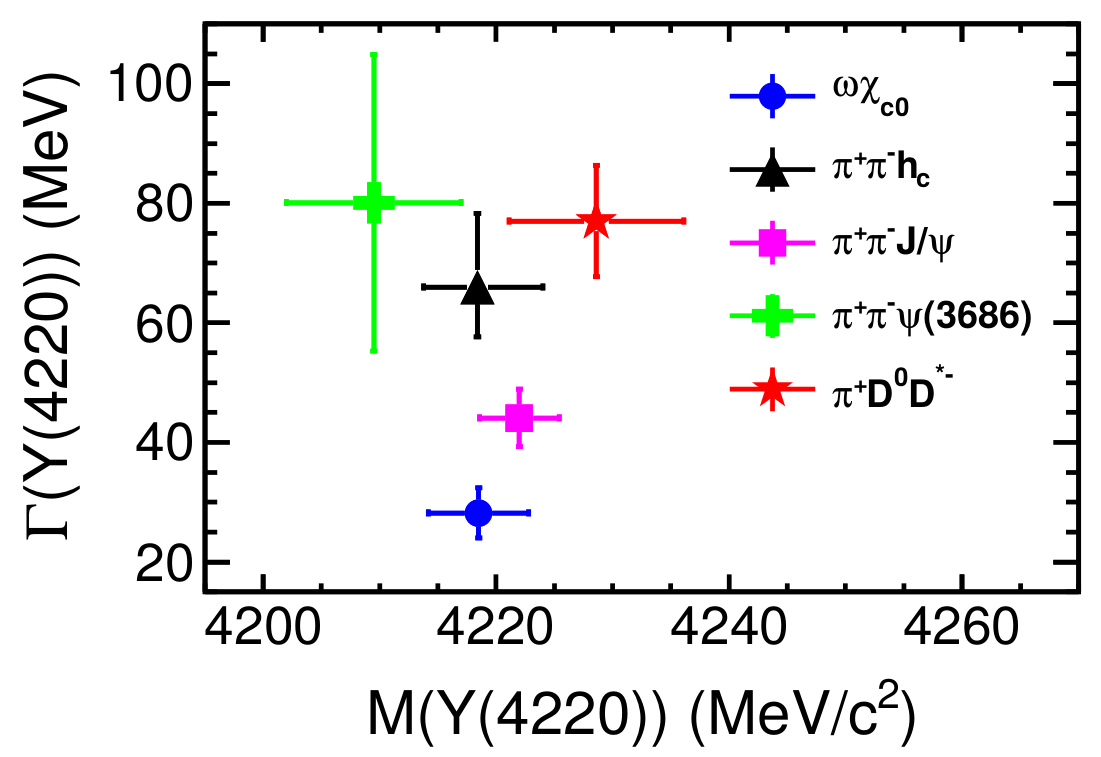}
\caption{$Y(4220)$ mass and width from different results}
\label{fig:fig6}
\end{subfigure}
\caption{Summary of recent results $Y(4220)$ using BESIII data}
\end{figure}

\subsubsection*{$e^+e^- \rightarrow  \pi^+ \pi^- D^0 \bar{D}^0$ \hspace{1cm} $e^+e^- \rightarrow  \pi^+ \pi^- D^+ D^-$}

A prediction of the molecular nature of the $Y(4220)$ can be tested by measuring the $e^+ e^- \rightarrow D_1(2420)\bar{D}$ cross section. \cite{Cleven:2013mka} predicted that a rapid rise of the cross section is an indicator of the molecule nature of this state. BESIII contribution was given by the investigation of  the $D_1(2420)$ meson decay into $D \pi$ pairs \cite{Ablikim:2019faj}. 
Here some of the cross section preliminary results performed in 15 different center of mass energy between $4.09$ and $4.60~\mathrm{GeV}$ are presented.
Three decay mode were under investigation $D_1(2420) \rightarrow D^0 \pi^+ \pi^-$, $D_1(2420) \rightarrow D^{*-} \pi$ and $D_1(2420) \rightarrow D^+ \pi^+ \pi^-$. A clear observation of the first decay mode is found at $4.42~\mathrm{GeV}$, no evidence of the second type of process is found, while evidence of the third process are found both at $4.36$ and $4.60~\mathrm{GeV}$.


The simultaneous fit of the three line shape of the $D_1(2420)\bar{D}$ cross section presented in Fig~\ref{fig:fig7} does not  show the expected sharp rise: those data seems to disfavour the molecular interpretation for the $Y(4220)$ state.

\begin{figure}[htb]
\centering
\includegraphics[height=2in]{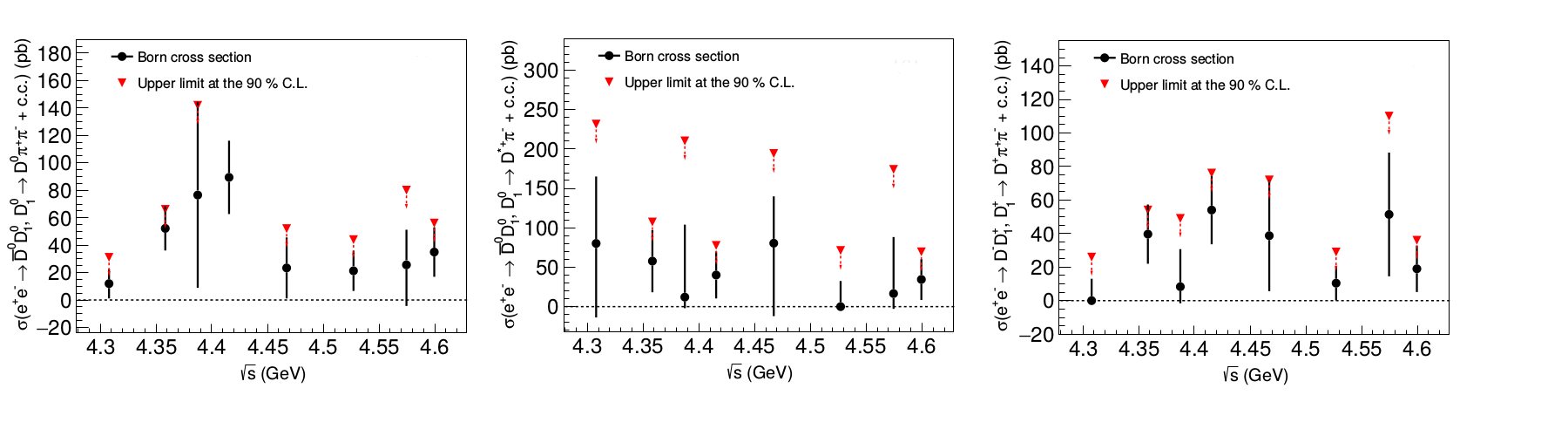}
\caption{Dressed cross section measurements. (Left) $D_1(2420) \rightarrow D^0 \pi^+ \pi^-$; (Center)$D_1(2420) \rightarrow D^{*-} \pi$; (Right) $D_1(2420) \rightarrow D^+ \pi^+ \pi^-$ }
\label{fig:fig7}
\end{figure}

\section{Conclusions}

Thanks to the possibility to directly produce the vector charmonium(-like) states with a very high luminosity, BESIII is playing a central role in the study of the nature of the XYZ states. Thanks to the great precision of the BESIII data, it is possible to extract the total measured branching ratio of $X(3872)$ decays up to now. Combining the most recent BESIII results with the ones from other experiments is possible to declare that the branching ratio of unknown decays is $ \left( 31.9^{+18.1}_{-31.5}\right)\%$ \cite{Li:2019kpj}. Moreover, thanks to the great statistics, other investigations can also be performed by the experiment to study the connections between these states and the conventional ones \cite{Yuan:2018inv}. \\
BESIII is going under a large upgrade. From the accelerator point of view the maximum center of mass energy will be extended up to $4.7~\mathrm{GeV}$ next year and up to $4.9~\mathrm{GeV}$ in the future. From the detector side a more performing inner tracker will be installed to help increase the background rejection and have better secondary vertex reconstruction. \\
This will allow BESIII to maintain its prime role in the search and discovery of exotics states in the oper charm region.

\end{document}